\documentclass[prl,aps,twocolumn,showpacs]{revtex4}
\usepackage{graphicx}
\begin{document}
\title{Quantum dynamics of bosons in a double-well potential: Josephson oscillations, self-trapping and ultralong tunneling
times}
\author{A. N. Salgueiro and A. F. R. de Toledo
Piza} \affiliation{Instituto de F\'{\i}sica, Universidade de S\~ao
Paulo, CP 66318 CEP 05389-970, S\~ao Paulo, SP, Brazil}
\author{G. B. Lemos, R. Drumond and M. C. Nemes}
\affiliation{Departamento de F\'{\i}sica, Instituto de Ci\^{e}ncias
Exatas, Universidade Federal de Minas Gerais, CP 702, CEP 30161-970,
Belo Horizonte, Minas Gerais, Brazil}
\author{M. Weidem\"uller}
\affiliation{Physikalisches Institut, Albert-Ludwigs-Universit\"at
Freiburg, 79104 Freiburg, Germany}

\begin{abstract}
The dynamics of the population imbalance of bosons in a
double-well potential is investigated from the point of view of
many-body quantum mechanics in the framework of the two-mode
model. For small initial population imbalances, coherent
superpositions of almost equally spaced energy eigenstates lead to
Josephson oscillations. The suppression of tunneling at population
imbalance beyond a critical value is related to a high
concentration of initial state population in the region of the
energy spectrum with quasi-degenerate doublets resulting in
imbalance oscillations with a very small amplitude. For
unaccessible long times, however, the system recovers the regime
of Josephson oscillations.
\end{abstract}


\maketitle

The understanding of many-body quantum systems from the
theoretical and experimental points of view has undergone a
considerable development during the past decade. Unifying concepts
of several branches of physics are under development, creating an
interdisciplinary scenario for the understanding of quantum
mechanical paradigms. One of the simplest many-body systems to be
realized experimentally and studied theoretically are ultracold
bosons in a double-well potential. This system is very rich
exhibiting a great variety of quantum phenomena such as
interference~\cite{ketterle},
tunneling/self-trapping~\cite{wall,smerzi1,smerzi,ober,wall2,zwerger},
entanglement of macroscopic superpositions~\cite{cirac}. Lately
this system has been extensively studied, especially after the
implementation of several experiments in the area. The usual
theoretical approach to weakly interacting Bose-Einstein
condensates (BECs)  is the mean-field approximation, a nonlinear
Gross-Pitaevski
equation~\cite{reinhardt,reinhardt3,wall,MF,GP,bergeman,piza,QPM,Bethe},
which has proven very adequate in explaining a wide variety of
experiments.

More recently, the dynamics of population distribution between two
or more wells of an optical lattice have been experimentally
investigated. In particular, Josephson oscillations have been
observed in a 1D optical lattice~\cite{kasevich,inguscio} and
recently the density distribution of tunneling $^{87}Rb$ particles
is directly observed~\cite{ober}. In this experiment, initial
population differences between the left and right well components
are controlled by loading the BEC into an asymmetric double-well
potential. The Josephson dynamics is initiated at $t=0$ by
non-adiabatically changing the potential to a symmetric
double-well. When the initial population imbalance is below a
critical value, the system presents Josephson oscillations between
the two sides of the well. However, above this critical value
tunneling is not observed. Based on a mean field treatment, this
is usually attributed to macroscopic self-trapping. In the present
work, we discuss an alternative approach to this system based on
exact numerical solutions of the two-mode Bose-Hubbard Hamiltonian
\cite{bh2}:
\begin{equation}
\label{um}
H=-\frac{J}{2}(a_1^{\dagger}a_2+a_2^{\dagger}a_1)+\frac{U}{2}(n_1(n_1-1)+n_2(n_2-1))
+\frac{\delta}{2} (n_1-n_2)
\end{equation}

\noindent where $a_i^{\dagger}$ and $a_{i}$ are, respectively, the
creation and annihilation operators of a boson in the $i$-th site,
$n_i$ the occupation number in the $i$-th site, $U$ is the
two-particle on-site interaction, $\delta$ is the trap depth
difference between the two wells, $J$ is the tunneling matrix
element between adjacent sites $i$,$j$. which characterizes the
strength of the tunneling term.

\begin{figure}
\centering
\includegraphics[width=\columnwidth]{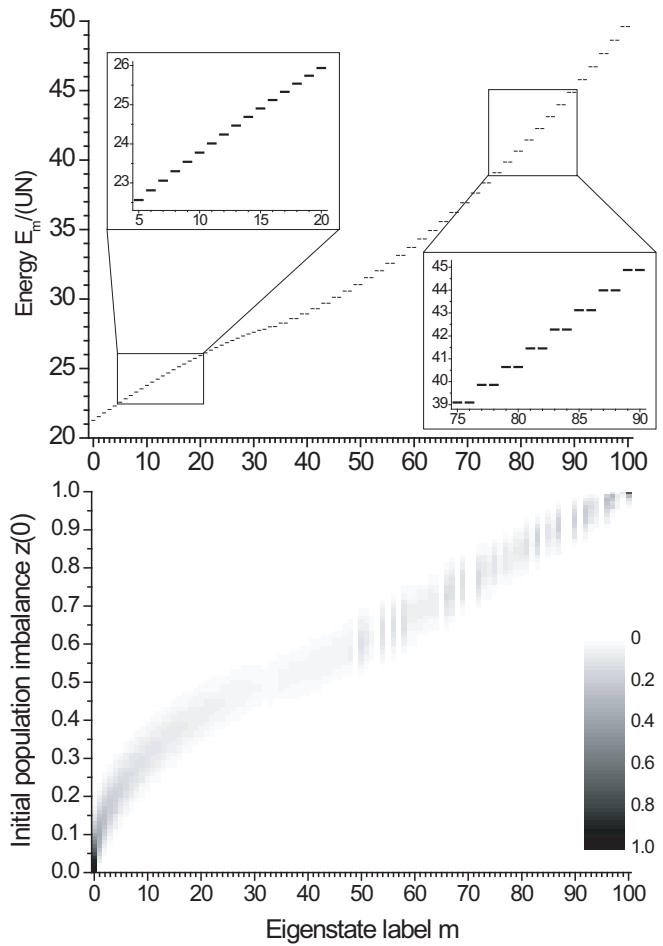}
\caption{Upper graph: Hamiltonian energy spectrum for 100 bosons
in a symmetric double-well potential for $J/U=3.333$ corresponding
to $\Lambda=15$. Lower graph: Density plot of the occupation
probability of symmetric Hamiltonian eigenstates $|m\rangle$ as a
function of the initial population imbalances $z(t=0)$ for the
same parameters as the upper graph. The gray scale gives the
population of the corresponding eigenstate. Below a critical value
$z_c\approx0.5$, only states with almost equal energy spacing are
populated, while above $z_c$ the quasi-degenerate doublet states
are occupied.}
\label{espectro10}
\end{figure}

In this Letter we present a different view on 
the experimental non-observation of tunneling for initial population
imbalances larger than a critical value $z_c$. The dynamics of the
system is related to the properties of the energy spectrum and to
the distribution of populations and coherences of the initial
wave-function. It is important to notice that this model displays
many-body features such as highly correlated many-body eigenstates.
In these eigenstates the number of atoms in each well is not defined
but features correlated fluctuations. According to the present
model, for certain range of the parameter $J/U$ the spectrum of the
many-body Hamiltonian presents a region with nearly degenerate
doublets, as shown in Ref.~\cite{wall} and Fig.~\ref{espectro10}.
For sufficiently large initial population imbalance, mainly doublets
will participate in the tunneling process. The tunneling period will
be approximately given by $t\sim \hbar/\Delta E_{doublets}$ where
$\Delta E_{\rm{doublet}}$ stands for the small inner energy
difference of the doublets. This energy splitting as $\Delta
E_{\rm{doublet}}/U=[2N(J/U)^N]/(N-1)!$ (calculated
via the doorway method of Ref~\cite{doorway}).  
For example, for $N=100$ particles, $J/U=0.333$, $\Delta
E_{\rm{doublet}}/U\sim 3.76\times10^{-202}\rightarrow 0$. This
energy splitting is usually very small and hence will result in
periods which are much longer than the time window of the
experiments. In the experimental situation of Ref.~\cite{ober}, only
much shorter times are observed which is far from the time the
tunneling actually happens, giving the idea of "self-trapping". This
short time regime is characterized by oscillations with small
amplitude which are related to the small coherence between the two
pairs of adjacent quasi-degenerate doublets
of the spectrum. For this case, the oscillation period is given by
the inverse of the energy splitting of adjacent quasi-degenerate
doublet, e.g., $t=\hbar/\Delta E$, with $\Delta E/U\simeq (N-1)$.
For this reason, the interpretation of the non-observation of
tunneling in the model is radically different from the one coming
from the Gross-Pitaevski equation, attributed to the non linearity
of the mean-field approach \cite{smerzi}. The same analysis can be
extended to the improved two-mode model developed in Ref.
\cite{bergeman}.






To illustrate our ideas, we study a hundred particles in a
double-well potential. The upper graphic in Fig.~\ref{espectro10}
shows the energy spectrum for the symmetric case at $J/U=3.333$ and
$N=100$, which corresponds to the parameter
$\Lambda\equiv\frac{NU}{2J}=15$ used in the experimental situation
of Ref.~\cite{ober}. The spectrum is composed of two regions: one
region of almost equidistant energy levels and one region consisting
of quasi-degenerate doublets. (see also Ref.~\cite{wall}). The
dynamics of population in these two regions differs vastly. It has
been shown in Ref.~\cite{reinhardt} that the eigenstates of the
system confirm the physical pendulum characteristics of the
eigenstates. The ground state is a minimum uncertainty wave packet
in both number and phase which is centered at the origin. The
harmonic-oscillator-like low-lying excited states are the analog of
pendulum librations, and the higher-lying cat like states are the
analog of pendulum rotor motions, with a clear signature of the
quantum separatrix state where the libration and rotation states
separate. This separatrix divides the energy spectrum in two
regions, with or without quasi-degenerate doublets which can be
directly observed in the upper graph of Fig.~\ref{espectro10}. The
existence of these two distinct spectrum regions has a direct
consequence for the dynamics. 


The tunneling dynamics is characterized by the time evolution of the
population imbalance

\begin{eqnarray}
z(t)&&\equiv\langle
\psi(t)|(n_2-n_1)|\psi(t)\rangle/N\nonumber\\&&=\sum_{m}z_{mm}+2\sum_{m<m'}
z_{mm'} \cos[\frac{(E_{m}-E_{m'})t}{\hbar}] \label{imbalance}
\end{eqnarray}

\noindent where $|\psi(t)\rangle=\sum_m c_me^{-iE_m t} |m\rangle$
and $z_{mm'}=c_{m}^{*}c_{m'}\langle
E_{m}|(\hat{n}_1-\hat{n}_2)|E_{m'}\rangle/N$, with
$E_m$ and $|m\rangle$ being, respectively, the eigenvalues and
eigenstates of the hamiltonian in the symmetric case ($\delta=0$ in
Eq.~(\ref{um})), $c_m=(\langle m|\psi(0)\rangle)$ and
$|\psi(0)\rangle$ is the
initial many-body wavefunction. 
Following the experimental approach of Ref.~\cite{ober} the initial
condition is then prepared by non-adiabatically transferring the the
ground state of an asymmetric double-well potential $\delta\neq0$ to
the symmetric double-well. This corresponds to a projection of the
ground state wavefunction for the asymmetric Hamiltonian
$\delta\neq0$ to the basis $|m\rangle$. The initial state thus
critically depends on the value of the parameter $\delta$ which is
characterized by the initial population imbalance $z(0)$.

Two regions of the energy spectrum can be accessed choosing
appropriate initial conditions for the many-body wavefunction of the
system. In the lower graph of Fig.~\ref{espectro10} the occupation
probability ${|c_m|}^2$ of the eigenstates $|m\rangle$ as a function
of the initial population imbalance $z(0)$ is depicted. For small
$z(0)$ only the lowest few energy eigenstates are populated. The
spread of the distribution of populated states increases as $z(0)$
increases, but for $z(0)$ below a critical value $z_c\approx 0.5$
only energy eigenstates with nearly equal energy splitting are
occupied (see upper graph in figure~\ref{espectro10}).  As follows
from Eq.~\ref{imbalance}, coherences between these states will lead
to a macroscopic oscillation of the population imbalance at a
frequency given by the energy splitting. For larger values
$z(0)>z_c$ this scenario changes and only doublet states are
occupied. These states are characterized by superposition states in
the basis of states confined to the left and right side of the
double. As these two states are almost degenerate, coherences among
them do not contribute to the oscillatory term in
Eq.~\ref{imbalance}.

\begin{figure}
\includegraphics[width=\columnwidth]{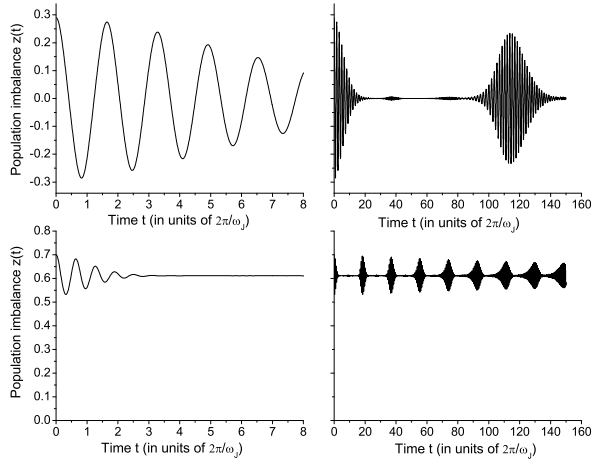}
\caption{Time evolution of the population imbalance 100 particles
for $J/U=3.333$ and initial conditions $z(0)=0.3$ (upper graph)
and $z(0)=0.7$ (lower graph).} \label{zt}
\end{figure}

The dynamical behavior of the system $z(t)$ crucially depends on the
competition of the two ingredients of $z_{n,m}$, the correlation
matrix of the initial state in the basis of the eigenstates of the
symmetric hamiltonian $c_m^{*}c_n$ and the correlation matrix of the
observable, the population imbalance in the basis of the eigenstates
of the symmetric hamiltonian $\langle E_{m}|(n_1-n_2)|E_{n}\rangle$.
The second term is responsible for the suppression of the population
oscillations of the first term. As the occupied eigenstates are
almost equally spaced for $z(0)<z_c$, the population imbalance
$z(t)$ oscillates around zero at a fundamental oscillation frequency
given by the mean energy spacing of the occupied states, which is
given by the plasma frequency $\omega_p=2J\sqrt{1+\Lambda}$, with
$\Lambda=(UN/2J)$ \cite{smerzi}. This behavior constitutes the
regime of Josephson oscillations as shown by the left graph of the
upper curves in Fig.~\ref{zt}. Small deviations from equal energy
spacing lead to a damping of the oscillation on a time scale
inversely proportional to the number of occupied states and the
corresponding differences in energy splitting. The oscillations
undergo revivals on time scales given by the inverse of the
frequency difference of adjacent eigenstates. The revivals can be
seen in the right graph of the upper curves in Fig.~\ref{zt}. For
larger values of $z(0)>z_c(0)$, where doublet states are occupied
with negligible coherences between adjacent doublet, one observes
two different time scales, characterized by a regime with small or
large oscillation amplitude. For the smaller time scales, the
population imbalance is locked to its initial value with small
residual oscillations as shown by the left graph of the lower curves
in Fig.~\ref{zt}. The frequency of these small oscillations is
determined by the inverse of the energy splitting of two
quasi-degenerate doublets, i.e. $\Delta E\simeq N U$. The latter
regime is commonly the "Self-Trapping" regime observed in
Ref.\cite{ober}. It should however be noted, that on larger time
scales $T\simeq\hbar/\Delta E_{\rm doublet}$, which are far beyond
experimental observation, the atoms still undergo collective
tunneling resulting in oscillatory behavior of the population
imbalance around zero, since the initial condition guarantees a high
occupation of the quasi-degenerate doublets, as one can see in the
lower graph of figure~\ref{espectro10}.

\begin{figure}
\centering
\includegraphics[width=0.7\columnwidth]{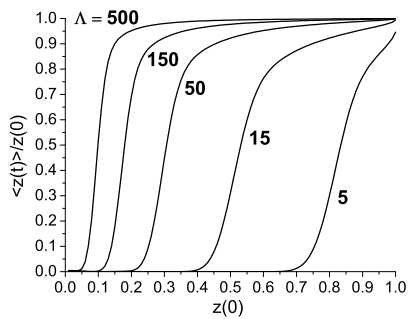}
\caption{Time average of $\bar{z(t)}/z(0)$ over 50 plasma periods
as a function of the initial population imbalance ($z(0)$) for 100
particles. Josephson oscillations result in $\bar{z(t)}=0$ while
$\overline{z(t)}\neq0$ indicates the regime of suppressed
tunneling. The parameter $\Lambda=NU/2J$ determines the critical
population imbalance separating these two regimes.} \label{meanzt}
\end{figure}

The time average of the population imbalance $\bar{z(t)}$ over 50
plasma periods $(2\pi/\omega_p)$ is shown in Fig~\ref{meanzt} for
different values of $\Lambda$ as a function of the initial
population imbalance $z(0)$ and gives the critical value $z_c$ for
which the system crosses from the Josephson regime (oscillations
around $z=0$) to a self-trapping regime. From Fig~\ref{meanzt} one
sees that as the parameter $\Lambda$ decreases, the critical value
$z_c$ increases for a fixed number of particles $N$. This can be
understood in terms of the structure of the symmetric hamiltonian
energy spectrum. The number of quasi-degenerate doublets increases
as the tunneling parameter $J/U$ decreases ($\Lambda$ increases). As
a consequence, the quasi-degenerate doublets have lower energies
which can be accessed by lower values of $z_c$, providing an
increase of the self-trapping region, since the region of high
distribution of the initial population imbalance in the
quasi-degenerate doublets also increases. The critical values $z_c$
of Fig~\ref{meanzt} follow very well the semiclassical prediction
$z_c\simeq 2\sqrt{1+\Lambda}/\Lambda$.

In spite of the obvious advantages of an exact approach, there are
several limitations to the model. The two mode assumption is an
approximation to the real situation. More restrictive, however, is
the hypothesis that the coefficients $J$ and $U$ remain constant in
time. This approximation is only valid when the many-body
interactions produce small modifications on the ground state
properties of the individual wells. This is true if the on-site
interaction energy is much smaller than the level spacing of the
external trap (see \cite{wall}). In this case the number of atoms
$N\ll \sqrt{\frac{\pi}{2}}\frac{r_0}{|a|}$, where $a$ is the
scattering length and $r_0$ is the position uncertainty in a
harmonic oscillator ground state. In the experiment with $^{87}Rb$
of ref.~\cite{ober,diss}, $N\ll 200$. However, qualitative behavior
is still observed.

We have investigated the dynamics of the imbalance population of
bosons in a double-well potential from the point of view of a
many-body Hamiltonian in the framework of the two-mode model.
Although the model is not realistic enough for large numbers of
particles ($N\geq200$), it points out a completely different
explanation for the suppression of tunneling. There are no
nonlinearities in this system and therefore the tunneling process is
explained solely in terms of initial conditions and spectral
properties of the many-body Hamiltonian. In this context, the
spectrum is divided in two regions, one of them consisting of
quasi-degenerate doublets. When the initial condition is small such
that $z(0)<z_c$, the occupation probability is larger in the lower
part of the energy spectrum with quasi-equidistant levels,
contributing with large oscillation amplitude with a plasma
frequency $\omega_p$. When the initial population imbalance is large
enough $z(0)>z_c$ only the quasi-degenerate doublets are appreciably
occupied. 
For small time range, the system oscillates with a very small
amplitude, whose oscillation period is given $t\simeq
\hbar/(U(N-1))$. However, for longer times, the system recovers
the Josephson oscillation behavior, with a large amplitude and
oscillatory period of $t\simeq\Delta E_{\rm{doublet}}/U=
(\frac{2N(J/U)^N\hbar}{(N-1)!})$ .In this case the tunneling time
given for a large number of particles is beyond observational
possibility. Hence, the "self-trapping" observed in the experiment
of ref.\cite{ober} is related to the behavior of
the imbalance population for short time. 
The same analysis can be extended to the improved two-mode model of
ref\cite{bergeman}, which includes  the terms reflecting the physics
of the transfer of atoms from one well to the other due to collision
excluded from the two-mode model,
since the energy spectrum of the improved model has the same
structure as the two-mode model, i.e. it is also divided into a
region with and without quasi-degenerate doublets. 
Therefore, the same qualitative physical picture would result by
this extension of the present model.

\begin{acknowledgments}
A.N.S. is supported by FAPESP and acknowledges M. Oberthaler for
discussions. M.C.N and R.D. thank CNPq for financial support.
G.B.L thanks CAPES for financial support. M.W acknowledges
financial support by DFG under WE $2661/1-3$ and DAAD in the
framework of the PROBRAL programme.
\end{acknowledgments}

\end{document}